\documentclass[prl,twocolumn,showpacs,preprintnumbers,amsmath,amssymb,superscriptaddress]{revtex4}

\usepackage{graphicx}
\usepackage{dcolumn}
\usepackage{bm}
\usepackage{color}

\begin{document}

\title{Thermodynamically Stable Blue Phases}

\author{F.~Castles}
\author{S.~M.~Morris}
\affiliation{Centre of Molecular Materials for Photonics and Electronics, Department of Engineering, University of Cambridge, 9 JJ Thomson Avenue, Cambridge CB3 0FA, United Kingdom}
\author{E.~M.~Terentjev}
\affiliation{Cavendish Laboratory, University of Cambridge, JJ Thomson Avenue, Cambridge CB3 0HE, United Kingdom}
\author{H.~J.~Coles}
\affiliation{Centre of Molecular Materials for Photonics and Electronics, Department of Engineering, University of Cambridge, 9 JJ Thomson Avenue, Cambridge CB3 0FA, United Kingdom}

\begin{abstract}
We show theoretically that flexoelectricity stabilizes blue phases in chiral liquid crystals. Induced internal polarization reduces the elastic energy cost of splay and bend deformations surrounding singular lines in the director field. The energy of regions of double twist is unchanged. This in turn reduces the free energy of the blue phase with respect to that of the chiral nematic phase, leading to stability over a wider temperature range. The theory explains the discovery of large temperature range blue phases in highly flexoelectric ``bimesogenic'' and ``bent-core'' materials, and predicts how this range may be increased further.
\end{abstract}


\maketitle

The blue phases (BPs) are liquid crystalline; they have orientational molecular order, yet flow like liquids. Up to three thermodynamically distinct BPs are observed upon cooling from the isotropic liquid: BPIII, BPII, and BPI respectively. BPIII is amorphous, in that it lacks long-range translational order, whereas BPII and BPI are periodic in three dimensions. They exhibit a frustrated structure consisting of a network of singular ``disclination'' lines in the field describing the average orientation of the molecules - the ``director'' field. The periodicity is typically of the order of the wavelength of visible light. This leads to vivid colored Bragg-like reflections, and a partial 3D photonic band gap. They are thus perhaps the only example of self-assembled 3D photonic crystals which, because of their fluidity, have optical properties that are readily switchable in an applied electric field \cite{wright,coles5,kitzerow,heppke,etchegoin}. Potential device applications include 3D lasers \cite{cao,morris2} and displays \cite{hisakado}.

Temperature stability is crucial to many applications; yet, the width of the entire BP region is only $\sim 1^\circ$C in typical liquid crystal (LC) materials. Polymer stabilization has been employed to widen this temperature range \cite{kikuchi}; however, the polymer network leads to restricted tunability and does not represent a true thermodynamic stabilization of the phase. BPs stable over a range of up to $50^\circ$C, including room temperature, were reported in 2005 \cite{nature}. The realization of polymer stabilized and nonpolymer stabilized BPs with wide temperature ranges has led to a resurgence of interest in the area. In the latter case a new theoretical challenge arose: to understand the mechanism of stabilization. In this Letter we demonstrate that it is a result of internal flexoelectric polarization. 

Flexoelectricity in LCs is a linear coupling between applied electric field and induced distortion \cite{meyer}. The inverse effect results in naturally distorted director structures, such as the BPs or LC colloids, having a flexoelecrically induced internal polarization field. This deformation-induced polarization has the effect of renormalizing the curvature elastic energy of director distortions. Recently, materials with nonsymmetric molecules have been developed with an unusually large flexoelectric response \cite{coles,harden,morris}. The stable BPs of Ref.~\cite{nature} were composed of ``bimesogenic'' LCs, for which very large flexoelectric coefficients have been reported \cite{coles}.

\begin{figure}
\includegraphics[width=0.45\textwidth]{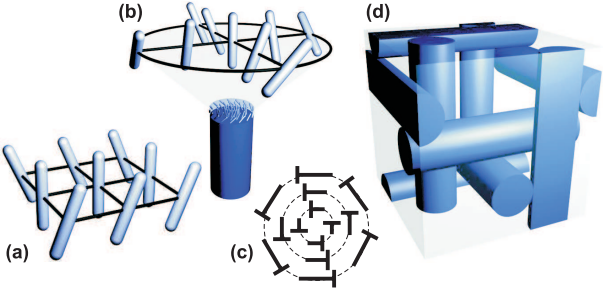}
\caption{\label{fig:dtc} (Color online) (a) Single twist in the chiral nematic
phase. (b) Regions of double twist may form double twist cylinders (DTCs). (c)
Cross section view of a DTC. (d) BPI is composed of a cubic arrangement of DTCs
with space group $O^{8-} (I4_132)$.}
\end{figure}

Upon cooling, the BPs usually undergo a transition to the helical cholesteric, or ``chiral nematic'' phase (denoted N*). The stability of the BPs is primarily a measure of their free energy with respect to that of the N*. The fact that the BPs can be stable compared to the defect-free N*, even in a narrow temperature range, is initially somewhat surprising: they contain a dense network of disclination lines that are energetically unfavorable due to high elastic distortion. Following Meiboom \textit{et~al.} \cite{meiboom, meiboom2} the stability can be understood by the fact that the N* is not \emph{locally} the lowest free energy configuration for chiral molecules. In the N*, the chiral molecules reduce their free energy by twisting with respect to each other along the helical axis, Fig.~\ref{fig:dtc}(a). However, molecules within pseudo planes perpendicular to that axis are still constrained to lie such that their average orientation is parallel to each other. A lower energy configuration is achievable in regions of ``double twist''. Such regions cannot extend indefinitely without topological defects, though they can extend parallel to the director, and a short way in directions perpendicular to this, to form cylinders of double twist, Figs.~\ref{fig:dtc}(b,c). BPII and BPI may be considered to be composed of these structures, called double twist cylinders (DTCs), arranged in regular cubic lattices, e.g., Fig.~\ref{fig:dtc}(d). The DTCs have free energy lower than the corresponding N*. In between the DTCs are a network of disclination lines of topological charge $s=-1/2$.  Because of the energy cost of elastic distortions, the free energy in the vicinity is higher than that of the corresponding N*. If the energy saving from the DTCs outweighs the energy cost due to the disclination lines at a given temperature, the BP will be stable with respect to the N* at that temperature \cite{meiboom, meiboom2}.
\begin{figure}
\includegraphics[width=0.45\textwidth]{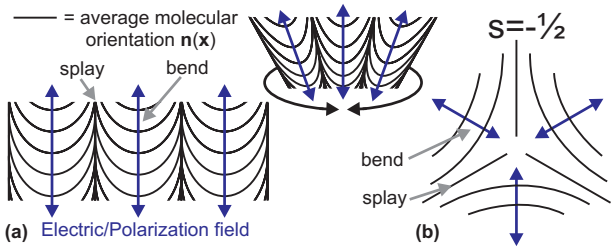}
\caption{\label{fig:sb} (Color online) (a) Flexoelectric materials may form a
characteristic splay-bend deformation of the director ${\bf n(x)}$ in an
external electric field  \cite{patel,meyer}. (b) The $s=-1/2$ disclination lines
are regions of alternating splay and bend. }
\end{figure}

It is interesting to note that the director profile of the $s=-1/2$ disclination is the splay-bend deformation, Fig.~\ref{fig:sb}(b), characteristic of flexoelectric materials in an external electric field \cite{meyer, patel}, rotated onto itself, Fig.~\ref{fig:sb}. Thus, one may intuitively expect that highly flexoelectric materials would form
such a distorted structure more readily. Yet, in the absence of external electric fields, a theory based on the local free energy density must be completely determined by the elastic coefficients. In this case, the stability of the BPs must be the result of a particularly favorable combination of elastic coefficients. By explicitly considering the energy of the internal
polarization field, its effect on renormalizing (in effect, reducing) the values of the corresponding elastic coefficients may be investigated \cite{helfrich,pikin}. To second order in spatial gradients of {\bf n}, and second order in induced electric displacement ${\bf D}$, the free energy density may be written in symbolic matrix notation:
\begin{equation}
f={\bf k_1^D} \bm \partial {\bf n}+\frac{1}{2}{\bf k_2^D} ( \bm \partial {\bf
n})^2 + {\bf e}\, \bm\epsilon^{-1} \bm\partial {\bf n}\, {\bf D} + \frac{1}{2}
\bm\epsilon^{-1} {\bf D}^2,
\end{equation}
where tensors ${\bf k_1^D}$ and ${\bf k_2^D}$ are elastic coefficients at constant ${\bf D}$, corresponding to the linear (chiral) and square of director gradients, $\bm\epsilon$ is the local dielectric permittivity tensor, and ${\bf e}$ represents the flexoelectric coefficients, which couple polarization to distortion. The dimensionality of boldface symbols is to be assumed, e.g, ${\bf k_2^D} ( \bm \partial {\bf n})^2$ may be written in a specific coordinate frame as $k^D_{2\,ijkl}\partial_i n_j \partial_k n_l$, where derivatives are with respect to position, and we use summation over repeated indices. The equilibrium value of ${\bf D}$ is found by minimization which, upon substitution back into $f$, gives
\begin{equation}
f={\bf k_1^D} \bm \partial {\bf n}+\frac{1}{2}\left({\bf k_2^D-{\bf
e}^2\bm\epsilon^{-1}}\right) ( \bm \partial {\bf n})^2.
\end{equation}
Imposing the locally uniaxial chiral symmetry of the N*, characterized by the point group $D_{\infty}$, and using the fact that ${\bf n}$ is a unit vector, $f$ may be written in the more traditional form \cite{oseen, frank}
\begin{equation} \label{eq:frank}
\begin{split}
f = \frac{1}{2}\left(K_{1}^D-\frac{e_{1}^2}{\epsilon_\parallel}\right) \left(
\bm\nabla \cdot \mathbf{n}\right)^2 + \frac{1}{2}K_2 \left( \mathbf n \cdot
\bm\nabla \times \mathbf{n} + q\right)^2 \\
+ \frac{1}{2}\left(K_3^D-\frac{e_{3}^2}{\epsilon_\perp} \right) \left(
\mathbf n \times \bm\nabla \times \mathbf{n} \right)^2 \\
+\frac{1}{2}\left( K_2+K_4\right)\bm\nabla\cdot\left[\left({\bf n }\cdot
\bm\nabla\right){\bf n}- {\bf n}\left(\bm\nabla\cdot{\bf n} \right)\right].
\end{split}
\end{equation}
$K_1$, $K_2$, and $K_3$ are the splay, twist, and bend Frank elastic coefficients, the reciprocal length scale $q$ determines the local chiral twisting power, and $\epsilon_\parallel$ and $\epsilon_\perp$ are the corresponding components of the dielectric permittivity. The combination $(K_2+K_4)$ is the ``saddle-splay'' elastic coefficient \cite{stewart}. $e_1$ and $e_3$ are the flexoelectric coefficients as defined by Meyer \cite{meyer}. We see that, for nonzero flexoelectric coefficients, $K_1$ and $K_3$ are reduced, while the twist coefficient $K_2$, and $K_4$ are unchanged (which is the consequence of splay and bend flexoelectric symmetry).

The reduction of Frank coefficients is $\sim 1\%$ \cite{pikin} in typical LCs, and therefore usually ignored. However, it will be more significant for the highly flexoelectric LCs now available (e.g. $\sim 20\%$, see below). The effect of this on the stability of the BPs may be understood as follows: the ``renormalization'' of $K_1$ and $K_3$ leads to a reduction of the energy cost of the disclination lines. Since $K_2$ and $K_4$ are unchanged, the energy saving due to double twist is unchanged. It has previously been shown, using numerical calculations, that the range of stability of the BPs is reduced as the ratio $(K_1+K_3)/K_2$ is increased \cite{meiboom2, alexander}.

In order to derive an analytic model to quantify the flexoelectric BP stabilization, we assume the following simplifications: the magnitude of the flexoelectric coefficients are equal $|e_1|=|e_2|\equiv \bar e$, the LC is dielectrically isotropic $\epsilon_\parallel=\epsilon_\perp\equiv\epsilon$, the unrenormalized elastic coefficients are equal $K_1^D=K_2=K_3^D\equiv K$,
and $K_4=0$. None of these assumptions alter the qualitative effect, but of course for a full quantitative agreement one would need to take into account particular different values of these coefficients in particular materials. Further, we ignore the effects of induced space charge, and nonlocal electric field effects ~\cite{dozov}. The effective splay and bend elastic coefficients are now equal to $rK$ where $r$ is the single renormalization factor from Eq.~(\ref{eq:frank}) 
\begin{equation} \label{eq:r} 
r=1-\bar e^2/(K\epsilon).
\end{equation}

Following Meiboom \textit{et~al.} \cite{meiboom,meiboom3}, the stability of the BPs may be investigated by considering the energy of a single $s=-1/2$ disclination line, required to topologically match neighboring regions of DTCs. The strength of this approach is that it leads to an analytic expression for the temperature range of stability, though there
may be limitations due to the inherent simplifying assumptions. The director field is approximately given by ${\bf n} = \left(\cos\left(s\phi \right), \sin\left(s\phi \right),0 \right)$, where $\phi$ is the polar angle in local cylindrical coordinates. Three contributions to the free energy per unit length are required. (1) The thermodynamic free energy cost of the core of the defect. This is proportional to the cross-sectional area of the core $F_\textrm{core}= \alpha\pi R^2$, where $R$ is the core radius. $\alpha$ is the energy density of the core, which is a function of the actual temperature relative to the isotropic transition temperature. (2) The elastic energy cost of splay and bend deformations surrounding the disclination, which is dependent on the renormalized coefficients $K_1$ and $K_3$. Including the flexoelectric renormalization, this energy is $F_\textrm{sb}=(1/4)\pi\,r\,K \ln\left(R_\textrm{max}/R\right)$ per unit length of disclination. Here, $R_\textrm{max}$ is the outer radius of the disclination line and is of the order of the distance between disclination lines in the BP. (3) The energy saving due to double twist. This is encoded through the volume integral of the splay-bend term in Eq.~(\ref{eq:frank}), converted to a surface integral over the surface of the core. Note that this term is dependent on the unrenormalized Frank coefficients $K_2$ and $K_4$, and can be shown in the one-constant approximation to be given by $F_\textrm{twist}=- \frac{1}{2}\pi K$ per unit length of a DTC \cite{meiboom}. The total energy $F_\textrm{core}+F_\textrm{twist}+F_\textrm{sb}$ is
\begin{equation}
F= \alpha\pi R^2 - \frac{1}{2}\pi K + \frac{1}{4} B\pi r
K\ln\left(R_\textrm{max}/R \right).
\end{equation}
A parameter $B$ is included to compensate for the inaccuracy of the analytic expression. Numerical calculations \cite{meiboom3, meiboom2} have estimated that $B\approx0.5$, a value that we now adopt. As a first approximation, the coefficient $\alpha$ is linear in temperature $T$, i.e., $\alpha = a(T_1-T)\equiv a\Delta T$ for $T<T_1$, where $T_1$ is the isotropic transition temperature and $a$ may be estimated from the latent heat of the isotropic-N* transition \cite{meiboom2}. The BP will be stable with respect to the N* when $F$ is negative. The core radius $R$ is determined by minimization of $F$, giving the ratio $R_c = \frac{1}{4} \sqrt{rK/\alpha}$, which is essentially the nematic correlation length, reduced by the renormalization factor $r$. Substituting this back into the free energy, the range of stability becomes
\begin{equation}  \label{eq:deltat2}
\Delta T=  A\, r \exp\left[(8/r)-1\right],
\end{equation}
where $A=K/\left(16aR_{\textrm{max}}^2 \right)$.

It is instructive to calculate the temperature range predicted by Eq.~(\ref{eq:deltat2}) for typical, experimentally determined, material parameters. Care must be taken: do the experimental values represent the ``bare'' or the renormalized elastic coefficients? In general, the values of the elastic coefficients will depend on the conditions of measurement, and be related \cite{ikeda} according to 
\begin{eqnarray} \label{eq:renorm}
K_{1}^E &=&K_1^D-e_1^2/\epsilon_\parallel=K_{1}^P-e_1^2/\chi_\parallel,  \\
K_{3}^E &=&K_3^D-e_3^2/\epsilon_\perp= K_{3}^P-e_3^2/\chi_\perp, \nonumber \\
K_2^E   &=&K_2^D=K_2^P, \quad\textrm{and}\quad K_4^E =K_4^D=K_4^P.\nonumber
\end{eqnarray}
Superscripts $E$, $D$, and $P$ denote measurements at constant electric field, displacement, and polarization respectively. Most experiments that measure these parameters employ the application of an alternating electric field, in which case constant-$D$ coefficients are measured \footnote{A further complication arises concerning the thermal conditions of measurement. For phase transitions such as those discussed here, isothermal coefficients will be relevant. In experiments using high-frequency alternating fields, adiabatic coefficients will generally be measured. In this paper we ignore the possible discrepancy.}. The ratio $|e_1-e_3|/K$ can be measured via ``flexoelectro-optic effect'' experiments, also under the application of high-frequency electric fields \cite{patel,coles}. Since $e_1$ and $e_3$ must generally be of opposite sign for a periodic splay-bend pattern to be formed \cite{meyer}, we assume this results in the ratio $2\bar e/K$ in our approximation \footnote{If $e_1$ and $e_3$ are of the same sign, the measured value of $|e_1-e_3|/K$ is a lower bound on $2\bar e/K$}. When combined with Fr\'eedericksz transition \cite{stewart} measurements, $\bar e$, $K$ and $\epsilon$ may be individually
determined.

Consider the typical values $a=8~\textrm{kJ}\,\textrm{K}^{-1}\,\textrm{m}^{-3}$, $R_\textrm{max}=100~$nm \cite{meiboom}, and experimental values for the bimesogenic material ``FFO11OCB'' ($|\bar e|=10.6~\textrm{pC\,m}^{-1}$, $K=(K_1+K_3)/2=8~\textrm{pN}$, $\epsilon=7.75\,\epsilon_0$, measured in \cite{morris}). These give a 20\% flexoelectric reduction of $K_1$ and $K_3$ ($r=0.8$). Equation~(\ref{eq:deltat2}) then predicts a temperature range of stability of $\Delta T = 43$~K, which is an order of magnitude greater than in usual chiral LCs, and is in approximate agreement with experiment on mixtures of such materials \cite{nature}. For comparison, if flexoelectricity is not taken into account in the above analysis, the predicted range is $\Delta T = 7$~K. Now consider the standard LC ``7CB,'' with experimental values
measured using the same methods ($|\bar e|=2.4~\textrm{pC\,m}^{-1}$, $K=(K_1+K_3)/2=4.9~\textrm{pN}$, $\epsilon=10.5\epsilon_0$
\cite{coles4,ratna}). This gives a flexoelectric renormalization factor r=0.99, and the predicted range of BP stability increases from $\Delta T = 4.2$~K to only $\Delta T = 4.6$~K. Again, this is comparable with experimental results.

The model also explains two observations in bent-core LCs: (1) Blue phases may be induced by doping with bent-core molecules \cite{nakata}. (2) Unusually stable BPs exist in bent-core LCs \cite{liao}. This may now be understood by the fact that, like bimesogens, bent-core LCs have unusually large flexoelectric properties \cite{harden}.

As $r$ is reduced further, $\Delta T$ can become very large (for example, larger than the transition temperature $T_1$ itself). This is not unphysical, rather it indicates that the BP is stable with respect to the N* at all temperatures below the transition from the isotropic. In this case the BP will transition directly to the lower-temperature smectic or crystalline phase, at which point the theory is cut off. The model may be refined to account for the nonlinear dependence of the thermodynamic free energy density of the disclination core on temperature far from the transition. The core energy $\alpha$ measures the difference between the ordered bulk and the disordered core. It scales with temperature as the (square of) the order parameter does, and at low temperatures becomes constant. To take this into account, we may instead let $\alpha = a T_1\left(1-\exp\left[T/T_1-1\right]\right)$ (or any another function that saturates in the desired manner).  In this case the predicted range of stability is greater still (46~K for FFO11OCB using the above values).

A numerical minimization of the free energy, such as that undertaken in Refs.~\cite{meiboom2, alexander}, but which further includes the flexoelectric renormalization, would be illuminating. In this way, the assumptions used to generate the analytic theory could be relaxed. In particular, it should be noted that our analysis of the disclination core is highly simplified, and the accuracy of the predictions may be improved by using a tensor order parameter that will account for the specific symmetry of the core. Nevertheless, our analytic model makes clear the strong dependence of the temperature range of stability on the flexoelectric coefficients, and the inherent approximations are somewhat validated through the above comparison with experiment. The theory makes two key predictions: (1) Further stabilization of the BPs can be achieved by ``molecular engineering" of more highly flexoelectric LCs. For example, by further exploiting the known relationships between molecular structure and flexoelectricity (see, e.g., Refs.~\cite{straley,coles}). (2) The dependence of the splay and bend elastic coefficients on the flexoelectric coefficients is large in certain materials, and should be accessible to experimental verification. Indeed, for highly flexoelectric materials, the flexoelectric properties may be accurately determined simply by measuring the elastic and dielectric properties under different conditions, according to Eqs.~(\ref{eq:renorm}).

We conclude that flexoelectricity strongly affects the energetics of distorted equilibrium structures. The internal polarization field affects all splayed and/or bent structures (since all LCs are, to some extent, flexoelectric), but is particularly relevant in highly flexoelectric LCs made of nonsymmetric molecules. This accounts for the existence of BPs that are stable over an unusually large temperature range.

This work was supported by Grants No. EP/D04894X/1 and No. EP/C537564/1.

\end{document}